%Paper: hep-th/9401165
%From: Jean-Bernard Zuber <zuber@amoco.saclay.cea.fr>
%Date: 31 Jan 94 15:38:31+0100

\input harvmac

\catcode`\@=11
\def\em@rk{\hbox{}}
\def\xeqn{\expandafter\xe@n}\def\xe@n(#1){#1}
\def\xeqna#1{\expandafter\xe@na#1}\def\xe@na\hbox#1{\xe@nap #1}
\def\xe@nap$(#1)${\hbox{$#1$}}
\def\eqns#1{(\e@ns #1{\hbox{}})}
\def\e@ns#1{\ifx\und@fined#1\message{eqnlabel \string#1 is undefined.}%
\xdef#1{(?.?)}\fi \edef\next{#1}\ifx\next\em@rk\def\next{}%
\else\ifx\next#1\xeqn#1\else\def\n@xt{#1}\ifx\n@xt\next#1\else\xeqna#1\fi
\fi\let\next=\e@ns\fi\next}
\catcode`\@=12
\lref\rBBIPZ{E. Br\'ezin, C. Itzykson, G. Parisi and J.-B. Zuber,
{\it Comm. Math. Phys.} 59 (1978) 35\semi Bessis, C. Itzykson, and J.-B.
Zuber, {\it Adv. Appl. Math.} 1 (1980) 109.}
\lref\rDSBKGM{M. Douglas and S. Shenker, {\it Nucl. Phys.} B335 (1990)
635\semi E. Br\'ezin and V. Kazakov, {\it Phys. Lett.} B236 (1990) 144\semi
D. Gross and A. Migdal, {\it Phys. Rev. Lett.} 64 (1990) 127;
{\it Nucl. Phys.} B340 (1990) 333.}%
\lref\rMehta{M.L. Mehta, {\it Random matrices, second edition}, ed. by Academic
Press inc (1991).}
\noblackbox
%\writedefs%\writestop
%\writetoc
%\hsize=162truemm\vsize=235truemm
%\voffset=-1.5truecm\hoffset=-0.5truecm
%\draftmode\baselineskip=16pt plus 2pt minus 1pt

\def\frac#1#2{{\textstyle{#1\over#2}}}

\def\tr{{\rm tr}\,}

\def\ud{\half}
\def\ee#1{{\rm e}^{^{\textstyle#1}}}
\def\d{{\rm d}}
\def\e{{\rm e}}

\def \l{\lambda}
\def \om{\omega}
\def \Om{\Omega}
\def \roc{{\cal O}}
\def \O{{\cal O}}
\def \s#1{\sqrt{(#1-a)(#1-b)}}

\Title{\vbox{\baselineskip12pt\hbox{SPhT/93-999}}}
{Large Random Matrices: Eigenvalue Distribution}

\vskip-\smallskipamount
\centerline{B. Eynard}

\bigskip{\baselineskip14pt
\centerline{Service de Physique Th\'eorique de Saclay}
\centerline{F-91191 Gif-sur-Yvette Cedex, FRANCE}
\centerline{Email: eynard@amoco.saclay.cea.fr}}
\bigskip

\footnote{}{*Laboratoire de la Direction des Sciences de la Mati\`ere
du Commissariat \`a l'Energie Atomique}

\vskip.2in

\centerline{Abstract:}
A recursive method is derived to calculate all eigenvalue correlation
functions of a random hermitian matrix in the large size limit, and after
smoothing of the short scale oscillations.
The property that the two-point function is universal, is recovered and
the three and four-point functions are given explicitly. One observes that
higher order correlation functions are linear combinations of universal
functions with coefficients depending on an increasing number of parameters of
the matrix distribution.

\vskip.3in
\Date{11/93, for Nuclear Physics B}
\lref\rBrZi{E. Br\'ezin and J. Zinn-Justin, {\it Phys. Lett.} B288 (1992)
54.}
\lref\rHINS{S. Higuchi, C.Itoi, S. Nishigaki, N. Sakai, {\it Nucl. Phys.} B318
(1993) 63; {\it
Nonlinear Renormalization Group Equation for Matrix Models}}
In a recent article \ref\rBrZe{E. Br\'ezin and A. Zee, {\it Nucl. Phys.}
B402 (1993); {\it Correlation functions in disordered systems}},
Br\'ezin and Zee have calculated explicitly correlation functions
of eigenvalues of a class of stochastic hermitian matrices of large size $N$.
They have found that some statistical properties of the eigenvalues are
universal in the large $N$ limit, and can thus be obtained from  the
correlation functions of the gaussian model.
More precisely, they have discovered that the two-point correlation function,
after smoothing of the short scale oscillations, is universal while all other
correlations vanish at the same order.

This property can be related to a renormalization group analysis
\refs{\rBrZi,\rHINS} which has shown that the gaussian model is a stable
fixed point in the large $N$ limit.

Their analysis is based on the, by now standard, method of orthogonal
polynomials. An essential ingredient in the final answer is a proposed ansatz
for an asymptotic form of the orthogonal polynomials $P_n$ in the limit
$N\to\infty$ and $N-n$ finite. In ref.~ \rBrZe\ the ansatz is  verified in
the case of even integrands, and only up to an unknown function.
\par
Here, we propose a direct proof of the ansatz, using a saddle point method,
which does not depend on the parity of the integrand,  and which
allows to determine the previously unknown function.
Moreover, using a completely different approach \ref\rACKM{J. Ambj\o rn, L.
Chekhov, C.F. Kristjansen, Yu Makeenko, {\it Nucl. Phys.} B404 (1993) 127; {\it
Matrix model
calculations beyond the spherical limit.}}, we present a recursive method to
evaluate all smoothed correlation functions at leading order. We give the
three and four-point functions explicitly.
\newsec{Correlation functions of eigenvalues}

Let us first explain the problem and recall the method used in ref.~\rBrZe\
to  explicitly evaluate  the eigenvalue correlation functions.\par

We consider $N\times N$ hermitian matrices $M$  with a probability
distribution of the form:
\eqn\eprob{{\cal P}(M)={1\over Z}\, \ee{-N\tr V(M)},}
where $V(M)$ is a polynomial, and $Z$ the normalization (i.e.\ the partition
function).  We want to derive the asymptotic form for $N$ large
of various eigenvalue correlation functions. All
can be derived from the correlation functions of the operator
$\O(\lambda)$:
\eqn\eOp{\O(\l)={1\over N}\tr \delta(\l-M)={1\over N}\sum_{i=1}^N
\delta\left(\l-\mu_i\right)}
(the $\mu_i$ being the eigenvalues of $M$).
Indeed, for any set of functions $(f_1,...,f_k)$, one has:
$${1\over N^k}\left< \tr f_1(M)\ldots\tr f_k(M)\right>=\int \d\!\l_1
f_1(\l_1)\ldots
\d\!\l_k f_k(\l_k)\, \left< \O(\l_1)\ldots \O(\l_k)\right>.$$
The correlation functions of the operator $\O(\l)$ can in turn be expressed in
terms
of the
partially integrated eigenvalue distributions
like $\rho(\l)$ the density of eigenvalues, which is the probability that
$\l$ belongs to the spectrum of $M$, $\rho_2(\l,\mu)$ the probability that $\l$
and $\mu$ are simultaneously  eigenvalues of $M$
and more generally $\rho_n(\l_1,\ldots,\l_n)$ the probability that
$\l_1$ ...$\l_n$ are simultaneously eigenvalues of $M$. For $\lambda_1\ne
\lambda_2 \ldots \ne \lambda_n$ (else some additional contact terms have to be
added)
we find
$$\left<\O(\lambda_1)\O(\lambda_2)\ldots \O(\lambda_n)\right>=
{1\over N^n}{N!\over (N-n)!} \rho_n(\l_1,\ldots,\l_n).$$

Actually the interesting functions are not directly the $\O(\lambda)$
correlation functions, but their connected parts. Indeed, at leading order,
when $N\to\infty$, we have the factorization property
$$\left<\O(\lambda_1)\O(\lambda_2)\ldots {\cal
O}(\lambda_n)\right> \sim \prod_{i=1,\ldots, n} \left<{\cal
O}(\lambda_i)\right>,$$
and thus no new information can be obtained from the complete
$n$-point function. The connected function which will be denoted
$$\left< \O(\l_1)\ldots \O(\l_n) \right>_{\rm
conn}=\roc_n(\l_1,\ldots,\l_n)
%\qquad {\rm for }\ \l_1\neq\l_2\ldots\neq \l_n%
$$
% ??? notation ???%
is only of
order $1/N^n$, and thus is  a subleading contribution.  The method of
orthogonal polynomials
allows to determine directly all these connected functions from only one
auxiliary function $\kappa(\lambda,\mu)=\kappa(\mu,\lambda)$ (see
refs.~\refs{\rBrZe,\rMehta}
or appendix~2 for details). For instance we have:
$$\eqalign{\roc_1(\l)&=\rho(\l)=\kappa(\l,\l) ,\cr
\roc_2(\l,\mu)&=-\kappa^2(\l,\mu) ,\cr
\roc_3(\l_1,\l_2,\l_3)&=2\kappa(\l_1,\l_2)\kappa(\l_2,\l_3)
\kappa(\l_3,\l_1),\cr
\roc_4(\l_1,\l_2,\l_3,\l_4)&=-2\kappa(\l_1,\l_2)\kappa(\l_2,\l_3)
\kappa(\l_3,\l_4)\kappa(\l_4,\l_1)
\cr&\quad -2\kappa(\l_1,\l_3)\kappa(\l_3,\l_2)\kappa(\l_2,\l_4)
\kappa(\l_4,\l_1) \cr
&\quad -2\kappa(\l_1,\l_2)\kappa(\l_2,\l_4)\kappa(\l_4,\l_3)
\kappa(\l_3,\l_1),\cr}$$
and analogous expressions for larger values of $n$:
$$ \roc_n=(-1)^{n+1}{1\over n}
\sum_{{\rm permutations}\
\sigma\,}\,\prod_{i=1}^n
\kappa\left(\l_{\sigma_i},\lambda_{\sigma_{i+1}}\right),$$
(each product of $\kappa$ appears exactly $n$ times in the sum and this
counting factor is cancelled by the $1/n$ factor). $\roc_n$ may also be
written:
$$ \roc_n=(-1)^{n+1}
\sum_{{\rm cyclic\, permutations}\
\sigma\,}\,\prod_{i=1}^n \kappa\left(\l_i,\lambda_{\sigma_i}\right)$$
The function $\kappa$ is
simply related to the polynomials $P_n$ orthogonal with respect to the measure
$\d\lambda\, \ee{-NV(\lambda)}$:
\eqn\eKdef{\kappa(\l,\mu) \propto {1\over N}
{P_N(\l)P_{N-1}(\mu)-P_{N-1}(\l)P_N(\mu)\over \l-\mu}
\exp[-(N/2)(V(\l)+V(\mu))] .}
The asymptotic evaluation of correlation functions is reduced
to an evaluation of the function $\kappa(\l,\mu)$ and thus of the orthogonal
polynomials $P_n(\l)$ in the peculiar limit: $N$ large, $N-n$ finite, and
$\l\in[a,b]$($[a,b]$ being the support of $\rho(\l)$).

The ansatz proposed in ref.~\rBrZe\ in the case of even potentials $V(M)$ (for
which $b=-a$)
was:
$$ P_n(\l)\propto\, \ee{N{V(\l)/ 2}}{1\over \sqrt{f(\l)}}\cos{\bigl(
N\zeta(\l)+(N-n)\varphi(\l)+\chi(\l)\bigr)},$$
where
$$\eqalign{\l= &a\cos\varphi\,, \cr
f(\l)& =a\sin\varphi\,, \cr
{\d\over \d\l}\zeta(\l)&=-\pi\rho(\l),\cr}$$
$\chi(\l)$ remaining undetermined, except in the gaussian and quartic
cases, for which: $\chi(\l)=\varphi/2 -\pi/4$.

We will prove below that this ansatz is still true for any $V$, and that
$\chi$ is always of the form $\chi=\varphi/2+{\rm const}$. The method
which leads to the proof is also interesting in itself because it uses some
general tools of the saddle point calculations of the one-matrix model
\refs{\rBBIPZ,\rDSBKGM}.
%This method could be possibly extended to other models:
%in the multi-matrix model, we can still write $\roc_n$ in terms of a
% kernel $\kappa$, which can be expressed by a Darboux--Christoffel theorem
% and we just have to determine the orthogonal polynomials $P_n$ and $Q_m$
%in the limit $N\to\infty$ and $N-n\sim 1$, $N-m\sim 1$. ???
%In the $O(n)$ model, there is no orthogonal polynomials, but the
%same methods apply for the determination of the resolvent.
%
\newsec{Orthogonal Polynomials}

Let us consider the set of polynomials ${P_n}$ ($n$ is the degree),
orthogonal with respect to the following scalar product:
\eqn\ePns{\left<P_n \cdot P_m\right> = \delta_{nm} = \int \d\l\, \ee{-N
V(\l)}\, P_n(\l)P_m(\l) .} %
Remarkably enough an explicit expression of these
orthogonal polynomials in terms of a hermitian matrix integral can be derived
(see appendix 1):
\eqn\ePnexact{P_n(\l)= Z_{n+1}^{-1/2}\int \d^{n^2}\! M\, \ee{-N \tr V(M)}
\det(\l -M),} where $M$ is here a $n\times n$ hermitian matrix, and
the normalization $Z_{n+1}$ the partition function corresponding to the
$n+1\times n+1$ matrix integral. We will use this expression
to evaluate $P_n$ in the relevant limit, i.e\ $N\gg 1$ and $N-n=O(1)$
by the steepest descent method. \par
Let us first introduce the matrix integral:
$$ Z(g,h,\l)=\int \d^{n^2}\! M\, \ee{-(n/g) \tr {\cal V}(M)}, $$
where
$$ {\cal V}(z)= V(z) -h\ln{(\l-z)}. $$
With these definitions $P_n$ is proportional to $Z(g=n/N,h=1/N,\l)$, and thus
we need $Z$ or equivalently the free energy $F_n=-n^{-2}\ln Z$ for $h$ and
$g-1$ small.
%consider arbitrary
%values of $g$ and $h$ to be able to take derivatives, and expand $F$ in
%powers  of $g-1$ and $h$.

As we are interested only in the $\l$ dependance of $F$, let us differentiate
$F$ with respect to $\l$:
$$ {\partial F\over \partial \l}=-{h\over g} \om(z=\l,g,h,\l)$$
where $\om(z)$ is the resolvent:
\eqn\eresolv{\om(z)={1\over n}\left< \tr{1\over z-M} \right> .}

Since we want the asymptotic expression of $ \om(\l)n^2 h/g$, we
need $\om$ up to the order $1/n$.
It is known from the random matrix theory \ref\rDGZ{For a review on matrix
model techniques see
also for example P. Di Francesco, P. Ginsparg and J. Zinn-Justin,
{\it 2D Gravity and Random Matrices} Saclay preprint T93/061, {\it Physics
Report} to appear.}, that $F_n$, and also $\om$ have
an expansion in powers of $1/n^2$ which is the topological expansion.
At order $1/n$, only the contribution of the sphere is required.
We thus replace $\om$ by its dominant contribution obtained by the saddle
point method.
\medskip
With this approximation, $\om$ may be written:
$$\om(z)={1\over n}\sum_{i=1,\ldots, n} {1\over z-\l_i}$$
where the $\l_i$ are the eigenvalues verifying the saddle point equation:

$${1\over g}{\cal V}'(\l_i)={2\over n}\sum_{j\neq i} {1\over \l_i-\l_j}\,,$$
i.e. they extremize the integrand of
$$ Z(g,h,\l)\propto \int \prod_{i=1}^{n} \ee{-N{\cal V}(\l_i)}d\!\l_i\,
 \prod_{i<j} (\l_i-\l_j)^2 \,.$$
In the large $n$ limit the $\l_i$ are distributed along an interval $[a,b]$
with a continuous density $\rho(\l)$. Then:
\eqn\eomrho{\om(z)=\int_a^b \d \mu\, \rho(\mu) {1\over z-\mu}}
and the saddle point equation becomes:
\eqn\elinear{{1\over g}{\cal V}'(\mu)=\om(\mu+i0)+\om(\mu-i0) \qquad {\rm
for}\quad\mu\in
[a,b]}
Note an important property of this equation: At $a$ and $b$ fixed it is linear
and therefore the derivatives of $\om$ with respect to $g$ or $h$ will
also satisfy a linear equation.

At leading order  we introduce the resolvent $\om_0(z)=\om(z,g=1,h=0)$, and
write:
$$ g\om(z,g,h,\l)=\om_0(z)+(g-1)\Om_g(z,\l)+ h \Om_h(z,\l) + O(1/n^2),$$
where we have defined the two functions:
$$\Om_g=\left. {\partial g\om\over \partial g}\right|_{g=0,h=1},\qquad
\Om_h=\left. {\partial g\om \over \partial h}\right|_{g=0,h=1}.$$

The function $\om_0(z)$ is the resolvent of the usual one-matrix model, and
from eqs.~\eomrho\ and \elinear\ we obtain:
$$\om_0(z\pm i0)={1\over 2} V'(z) \mp i\pi \rho(z) \qquad {\rm for} \quad z\in
[a,b].$$

As we noted above, the two functions $\Om_g$ and $\Om_h$ obey linear
equations, obtained by differentiation of the equation \elinear\ satisfied by
$\om$.
Following a method introduced in ref.~\ref\rEyZj{B. Eynard and
J. Zinn-Justin,  {\it Nucl. Phys.} B386 (1992) 558.} we can then easily
determine them from their analyticity properties, and the boundary
conditions.
 $\Om_g(z)$ verifies the linear equation:
$$\Om_g(z+i0)+\Om_g(z-i0)=0$$
and behaves as $1/z$ when $z\to\infty$, and as $1/\sqrt{z-a}\sqrt{z-b}$ near
the cut end-points $a$ and $b$, because $\om$ behaves as
$\sqrt{z-a}\sqrt{z-b}$.  These conditions determine $\Om_g$ uniquely:
$$\Om_g(z,\l)={1\over \sqrt{(z-a)(z-b)}}.$$
 The same method applies to $\Om_h$ which satisfies
$$\Om_h(z+i0)+\Om_h(z-i0)=1/(\l-z),$$
and behaves like $\Om_h\sim O(1/z^2)$ for $z$ large, $\Om_h\sim
1/\sqrt{z-a}\sqrt{z-b}$ near $a,b$, and is regular near $z=\l$. It follows:
$$\Om_h(z,\l)={1\over2}{1\over \sqrt{(z-a)(z-b)}}\left( 1-{
\sqrt{(z-a)(z-b)}-\sqrt{(\l-a)(\l-b)}\over z-\l} \right).$$
Then if we set $z=\l$:
$$\Om_h(z=\l)={1\over 2\sqrt{(\l-a)(\l-b)}}-{1\over 2}{\d\over
\d\l}\ln\sqrt{(\l-a)(b-\l)}.$$
\medskip
We now have the necessary ingredients to determine $\partial F/ \partial
\l$:
$$ -n^2{\partial F\over \partial \l}= N\om_0(z=\l) + (n-N) \Om_g(z=\l) +
\Om_h(z=\l) + O(1/N).$$

We still have to integrate all these terms with respect to $\l$.
In order to integrate $ \om_0= V'/2 -i\pi\rho$, we introduce a function
$\zeta(\lambda)$
$$\zeta(\lambda)=-\pi\int_a^\lambda\d\lambda'\rho(\lambda').$$

We also need the integral of $[(\l-a)(b-\l)]^{-1/2}$. For this purpose,
we parametrize $\l=\ud(a+b) -\ud(b-a)\cos\varphi$, so that the
integral of $\left[\ud(b-a)\sin\varphi\right]^{-1}$ is simply $\varphi$.
Finally, the result takes the form:
$$\eqalign{-n^2 F(\l\pm i0)&=\ud NV(\l)\pm iN\zeta(\l) \mp i(N-n)\varphi \pm
i\ud\varphi\cr&\quad -\ud\ln{ \sqrt{(\l-a)(b-\l)}} + {\rm const}\ .\cr}$$
Since $P_n$ is a polynomial, we have $P_n(\l)=\ud\left[ P_n(\l+i0)+P_n(\l-i0)
\right]$, and therefore:
\eqn\ePnasym{P_n(\l)=\sqrt{2\over\pi}\, \ee{NV(\lambda)/2}{1\over\sqrt{f(\l)}}
\cos{\left[
N\zeta(\l)-(N-n)\varphi(\l)+\ud\varphi(\l)+{\rm const} \right] } ,}
where $f(\l)= \sqrt{(\l-a)(b-\l)}$.
The constant factor $\sqrt{2/\pi}$ is fixed by the condition that
$\kappa(\l,\l)=\rho(\l)$.
In the case of even potentials $V$, parity considerations imply that the
arbitrary constant phase is ${\rm const}=-\pi/4$.
Indeed, we have $\zeta(a)=0$, $\zeta(b)=-\pi$, $\varphi(a)=0$,
$\varphi(b)=\pi$,  ( we
have $f(a)=f(b)=0$, but for $\epsilon$ a small positive number
:$f(a+\epsilon)/f(b-\epsilon)=1$) and thus:
$$P_n(b-\epsilon)/P_n(a+\epsilon)\sim (-1)^{n+1} \tan{({\rm const})}.$$
 For general potentials the constant phase
remains undetermined at this order, but we note that the general form of $P_n$
does not depend on the parity properties of~$V$.
\newsec{Connected Correlation Functions}

 From this asymptotic expansion of $P_n$ one can now derive the kernel
$\kappa(\l,\mu)$, and then the connected correlation functions, in the
large $N$ limit. The authors of ref.~\rBrZe\ have calculated some correlation
functions in two regimes: short range correlations $(\l_i-\l_j)\sim 1/N$, and
mesoscopic correlations \hbox{$(\l_i-\l_j)\gg 1/N$.}
Note that the polynomials $P_N$ and $P_{N-1}$ oscillate at a frequency of
order $N$ (which corresponds to the discrete spectrum of a matrix of size $N$
finite), and therefore, all the correlation functions will present such
oscillations.

In the short distance regime, these oscillations give the dominant behaviour,
and eq.~\ePnasym\ leads to:
$$\kappa(\l,\mu)\sim {\sin{[2\pi N (\l-\mu)\rho(\l)]}\over 2\pi N(\l-\mu)}$$
and all connected correlation functions follow.
Note that $\kappa$ being of order $1/N$, the connected microscopic $n$-point
function will be of order $1/N^n$. All this is studied in detail in
ref.~\rBrZe, and we will now concentrate our attention on the mesoscopic case.

In the regime $\l_i-\l_j\gg 1/N$, it is interesting to
consider smoothed functions, defined by averaging the fast oscillations.
For instance we find that:
\eqn\erhodeu{[ \roc_2(\l,\mu)]_{\rm smooth}={-1\over 2N^2\pi^2}{1\over
(\l-\mu)^2}{1-\cos\varphi \cos\psi\over \sin\varphi\sin\psi}\,,}
where
\eqn\epara{\l={a+b\over2}-{b-a\over2}\cos\varphi \,,\quad
 \mu={a+b\over2}-{b-a\over2}\cos\psi\,. }

Br\'ezin and Zee noted (ref.~\rBrZe) that the smoothed higher order
$n$-points correlation functions vanished identically at the order $1/N^n$ for
$n>2$.
Indeed, we will prove below that they are of order $1/N^{2(n-1)}$, and give a
recursive method to compute them with the help of loop-correlators.
\newsec{The $n$-Loop Correlation Functions}

Let us introduce the functions:
\eqn\edefloop{ \om_n(z_1,\ldots ,z_n)=N^{n-2}\left< \tr {1\over
z_1-M}\times\ldots\times\tr {1\over z_n-M}\right>_{\rm conn}}
They are related to the previous correlation functions by the relations:
\eqn \eomroc{\om_n(z_1,\ldots ,z_n)=N^{2n-2}\int \prod_{i=1}^n {\d\l_i\over
z_i-\l_i}
\,
\roc_n(\l_1,\ldots,\l_n)}

$\om_n$ is called the n-loop correlation function, because it is the
Laplace-transform of the partition function of a discrete random surface
limited by $n$ loops (the $z_i$ are conjugated to the lengths of the loops).
This remark allows to understand the topological origin of the factor
$N^{2n-2}$:indeed,

the Laplace-transform of the complete $n$-point correlation function which is
of order $1$, would be the partition function of every surface (not necessary
connected) with $n$ boundaries.
Each surface contributes with a topological weight $N^\chi$ where $\chi$ is
its Euler character. The leading term is the most disconnected one, with
$\chi=n$ (indeed, such a surface is made of $n$ discs, each of them having
$\chi=1$), while the connected term has $\chi=2-n$ (it is a sphere ($\chi=2$)
from which $n$ discs have been removed). Therefore, the relative contribution
of the connected part to the complete $n$-loop function is $N^{2-2n}$.

Remark that relation \eomroc\ can be inverted: $\om_n$ is analytical except
when some of the $z_i$ belong to the interval
$[a,b]$. $\roc_n$ can then be expressed in terms of the differences of $\om_n$
between opposite sides of the cut. For instance:
$$\eqalign{ \rho(\l)&={-1\over 2i\pi}(\om(\l+i0)-\om(\l-i0))\cr
\roc_2(\l,\mu)&={1\over
(2i\pi N)^2}(\om_2(\l+i0,\mu+i0)-\om_2(\l+i0,\mu-i0)-\om_2(\l-i0,\mu+i0) \cr
&\quad +\om_2(\l-i0,\mu-i0)).\cr}$$
and for general $n$:
\eqn\erocom{
\roc_n(\l_1, \dots,\l_n)={1\over N^{2n-2}}\left({-1\over 2i\pi}\right)^n
\sum_{\epsilon_i=\pm 1}
\left(-1\right)^{(\epsilon_1+\dots+\epsilon_n)}\, \om_n(\l_1+\epsilon_1
i0,\ldots,\l_n+\epsilon_n i0)}
Note that these functions are directly the smoothed correlation functions,
since we first compute $\om_n$ in the large $N$ limit at complex arguments
(which suppresses the oscillations), and then take the discontinuities along
the cut.

In order to compute $\om_n$, we consider the partition function:
$$ Z=\e^{N^2 F}=\int \d M\, \ee{-N\tr  V(M)}$$
with $V(z)=\sum_{k=1}^\infty g_k z^k/k$, and define the loop insertion
operator:
\eqn\eloopinsert{ {\delta\over \delta V(z)}=-\sum_{k=1}^\infty {k\over
z^{k+1}} {\del\over\del g_k}.}
Note that with this definition
\eqn\eVder{{\delta V(z')\over \delta V(z)}={1\over z-z'}-{1\over z}\,.}
The $\om_n$ are obtained from the free energy $F$ by the repeated action of
this operator:
\eqna\eloopn
$$\eqalignno{\om_n(z_1, \ldots, z_n)&={\delta\over\delta
V(z_1)}\cdots{\delta\over \delta V(z_n)}F \,,&\eloopn{a} \cr
&={\delta\over\delta V(z_n)}\om_{n-1}(z_1, \ldots, z_{n-1}).&\eloopn{b}\cr}$$
It is not necessary to calculate the free energy $F$, since we already  know
the one-loop function $\om(z)$.
We have already emphasized that $\om(z)$ satisfies a linear equation
(eq.~\elinear), and thus, all its derivatives satisfy the same linear
equation, with a different  l.h.s. . Then, analyticity properties, and
boundary conditions determine the form of $\om_n$.

The linear equation for $\om$ is:
\eqn\elooplin{ \om(\l+i0)+\om(\l-i0)=V'(\l).}
By a repeated action of the loop insertion operator, we obtain:
\eqn\eloopdeux{ \om_2(\l+i0,z)+\om_2(\l-i0,z)=-{1\over (z-\l)^2},}
and for $n>2$:
\eqn\enloop{ \om_n(\l+i0,z_2,\ldots,z_n)+\om_n(\l-i0,z_2,\ldots,z_n)=0\,.}
The function $\om(z)$ has the form
$$ \om(z)=\ud\left( V'(z)-M(z)\sqrt{(z-a)(z-b)}\right)$$
where $M(z)$ is a polynomial such that $\om(z)\sim 1/z$ when $z\to\infty$,
and therefore:
$$M(z)=\left({V'(z)\over \sqrt{(z-a)(z-b)}}\right)_+ \quad .$$
Then
$$\rho(\l)={1\over 2\pi}M(\l)\sqrt{(\l-a)(b-\l)}={1\over
2\pi}M(\l){b-a\over2}\sin\varphi\,,$$
where we have used the parametrization \epara.\par
The two-loop function is also completely determined by the linear equation
\eloopdeux\ and boundary conditions:
$$ \om_2(x,y)=-{1\over 4}{1\over (x-y)^2}\left(
2+{(x-y)^2-(x-a)(x-b)-(y-a)(y-b)\over
\sqrt{(x-a)(x-b)}\sqrt{(y-a)(y-b)}}\right)$$
and thus, in agreement with eq.~\erhodeu:
$$ \roc_2(\l,\mu)=-{1\over 2N^2\pi^2}{1\over
(\l-\mu)^2}{1-\cos\varphi\cos\psi\over \sin\varphi\sin\psi}.$$

The other loop functions all satisfy an homogeneous equation, and can be
written:
$$ \om_n(\l_1,\ldots,\l_n)=\left( \prod_{i=1}^n {1\over
\sin{\varphi_n}}\right)^{2n-3} W_n(\l_1,\ldots,\l_n),$$
with now $\l_n={a+b\over2}-{b-a\over2}\cos\varphi_n$, and
where the $W_n$ are some symmetric polynomials of degree less than $2n-5$ in
each $\l_i$, which are no longer determined by the boundary conditions.
It is necessary to directly use the recursion relation \eloopn{b}.
Since $\omega_2$ depends on the potential $V(M)$ only through $a$ and $b$,
we need the actions of loop-insertion operator on $a$ and $b$, for instance
$ {\delta a/\delta V(z)}$, ${\delta^2 a/\delta V(z)\,\delta V(z')}$.... For
this purpose, we introduce the following moments of the potential:
$$ M_k=-{1\over 2i\pi} \oint\d z\, {1\over (z-a)^k} {V'(z)\over
\sqrt{(z-a)(z-b)}}$$
$$ J_k=-{1\over 2i\pi} \oint\d z\, {1\over (z-b)^k} {V'(z)\over
\sqrt{(z-a)(z-b)}}$$
(the integration path turns clockwise around the cut $[a,b]$).

They are such that:
$$M(z)=\sum_k M_{k+1} (z-a)^k=\sum_k J_{k+1} (z-b)^k$$
The $M_k$ and $J_k$ are linearly related to the coefficients of the
potential $V$, and if $V$ is a polynomial of degree $v$, $M$ is of degree
$v-2$, and there are only $v-1$ independent coefficient among the $M_k$ and
$J_k$. Note also, that if $V$ is even, we have $M_k=-(-1)^k J_k$.

The cut end-points $a$ and $b$ depend on $V$ through the conditions that (see
ref.~\rDGZ~):
$$ \eqalign{ {1\over 2i\pi}\oint\d z\, {V'(z)\over \sqrt{(z-a)(z-b)}}
&=0\,,\cr
{1\over 2i\pi}\oint\d z\, {zV'(z)\over \sqrt{(z-a)(z-b)}} &=2\,.\cr }$$
It follows (using \eVder\ and performing the contour integrals):
$$\eqalign{{\delta a\over\delta V(z)}&={1\over M_1}{1\over z-a}{1\over
\sqrt{(z-a)(z-b)}}\,, \cr
{\delta b\over\delta V(z)}&={1\over J_1}{1\over z-b}{1\over
\sqrt{(z-a)(z-b)}}\,.\cr}$$
In order to determine the higher order derivatives of $a$ and $b$, we need to
differentiate the coefficients $M_k$ and $J_k$:
$$\eqalign{{\delta M_k\over \delta V(z)}=&{2k+1\over2} {\delta a\over\delta
V(z)}\left(M_{k+1}-{M_1\over (z-a)^k}\right)\cr
 &+{1\over2}{\delta b\over \delta V(z)}\left({J_1\over (b-a)^k}-{J_1\over
(z-a)^k}-\sum_{l=0}^{k-1}{M_{l+1}\over
(b-a)^{k-l}} \right) \cr }
$$
and analogous formulae are obtained for the $J_k$ by the exchange
$a\leftrightarrow b$.

With these tools, we can now determine the $\om_n$ recursively.
Let us for instance write $\om_3$:
$$\eqalign{\om_3(x,y,z)&={a-b\over 8\left(\s{x}\s{y}\s{z}\right)^3}\cr
&\quad\times\left( {1\over
M_1}(x-b)(y-b)(z-b)-{1\over J_1}(x-a)(y-a)(z-a)\right) \cr}$$
For $\omega_4$ we first define the polynomials
$$Q(x_i,a)=\prod_{i=1,4}\left(x_i-a\right).$$
With this notation
$$\eqalign{\om_4(x_i)&={-1 \over16 \left[Q(x_i,a)Q(x_i,b)\right]^{3/2}}
\left( -3(b-a){M_2\over M_1^3} Q(x_i,b)-3(a-b){J_2\over J_1^3}Q(x_i,a)
\right.\cr
&\quad +3{1\over M_1^2}Q(x_i,b)\left[
(b-a)\left(\sum\nolimits_{i=1}^4{1\over x_i-a}\right) -1\right]\cr
&\quad +3{1\over J_1^2}Q(x_i,a)\left[
(a-b)\left(\sum\nolimits_{i=1}^4{1\over x_i-b}\right) -1\right]\cr
& \left.\quad+{1\over M_1 J_1} \left[(x_1-a)(x_2-a)(x_3-b)(x_4-b)+\ {\rm
5\ terms}  \right] \right),\cr } $$
where the last additional terms symmetrize in the four variables.
%& -16\om_4(x,y,z,t)=\cr
%& -3(b-a)\left( M_2 M_1 a_x a_y a_z a_t- J_2 J_1 b_x b_y b_z b_t \right) \cr
%&+3 M_1^2 a_x a_y a_z a_t \left[ (b-a)\left( {1\over x-a}+{1\over
%y-a}+{1\over z-a}+{1\over t-a}\right)-1\right]\cr
%&+3 J_1^2 b_x b_y b_z b_t \left[ (a-b)\left( {1\over x-b}+{1\over
%y-b}+{1\over z-b}+{1\over t-b}\right)-1\right]\cr
%&+ M_1 J_1 \left[ a_x a_y b_z b_t+ a_x b_y a_z b_t+ b_x a_y a_z b_t\right.
%&+ \left. a_x b_y b_z a_t+ b_x a_y b_z a_t+ b_x b_y a_z a_t\right] \cr } $$
The connected functions $\roc_n$, are then obtained by \erocom\ and \enloop,
and they are simply given by:
$$\roc_n(\l_1,\dots,\l_n)={1\over N^{2n-2}}\left( {-1\over i\pi}\right)^n
\om_n(\l_1+i0,\dots,\l_n+i0).$$

{\it Universality}. We observe that the only universal features of ${\cal
O}_1(\lambda)=\rho(\lambda)$ are the square-root singularities at the edge of
the distribution, otherwise the function is potential dependent.
%
%The density $\rho(\l)$ is not universal at all, since it involves the
%complete
%$M(\l)$ function, i.e. all the momenta $M_k$ and $J_k$. In other words, two
%potentials
%
Instead the two-point correlation function has a universal form, and can thus
be calculated from the gaussian model.

The $n$-point smoothed correlation functions with $n\geq 3$ can be calculated
recursively by a systematic method described above. The main property is that
they consist in a sum of a finite number of universal functions and involve
only the first $n-2$ moments of the potential. That means for instance, that
two potentials $V$ and $V^*$ induce the same three-point function as soon as
they yield  the same $M_1$ and $J_1$ coefficients, but
they don't need to be identical. However, the $n$-point correlation function
is no longer given by a gaussian model, since it is of order $1/N^{2n-2}$.
Perturbative corrections to the gaussian model have to be considered.

Note that the determination of correlation functions allows an evaluation
of the moments $M_k$, in a case where the potential $V$ is unknown.
Let us emphasize that these moments play an important role in the study of
critical points. Since
$$\rho(\l)={1\over 2\pi}M(\l)\sqrt{(\l-a)(b-\l)}$$
we see that if some moments vanish, the behaviour of $\rho$ near the
end-points is no longer a square-root.  When the $m$ first $M_k$ vanish,
one finds $\rho\sim (\l-a)^{m+1/2}$ which corresponds to the $m^{\rm th}$
multicritical point of the one-matrix model of 2D gravity.
\refs{\rDSBKGM,\rDGZ}.

\newsec{Conclusions}

In this article we have recovered, by a completely different method, the
results of ref.~\rBrZe concerning the two-point eigenvalue correlation
function of a random hermitian matrix in the limit in which the size $N$ of
the matrix becomes large. Our method has allowed us to generalize the results
to new distributions. In addition, we have established a recursion relation
between successive $n$-point correlation functions at leading order for $N$
large.

Br\'ezin and Zee in ref.~\rBrZe have shown that the two-point function is
universal, and therefore identical to the function of the gaussian matrix
model. The gaussian model is the fixed point of a renormalization group
\refs{\rBrZi,\rHINS} and a direct RG analysis should be performed to put this
result in perspective.

In the same way, Br\'ezin and Zee have shown that higher correlation functions
vanish at leading order. The contributions we have calculated here should be
considered as corrections to the leading scaling behaviour. The explicit
expressions we have obtained show that they depend now on successive moments
of the potential, indicating an implicit classification of the deviations from
the gaussian model in terms of their irrelevance for $N$ large.
Here also it would be interesting to confirm the qualitative aspects of these
results by a direct RG analysis.

% Conclure, dire plus
%----------------------------------------------------------------%
\appendix{1}{Orthogonal polynomials $P_n$: An explicit expression}

Let us show that the orthogonal polynomials $P_n$ defined by the orthogonality
condition:
$$ \left<P_n \cdot P_m\right> = \delta_{nm} = \int \d \l\, \ee{-N V(\l)}\,
P_n(\l)P_m(\l) .$$
are given, up to a normalization, by equation \ePnexact:
$$ P_n(\l)\propto \int \d^{n^2} M\, \ee{-N \tr V(M)} \det (\l -M).$$
First, this integral clearly yields a polynomial of degree $n$ in $\l$.
Let us then verify the orthogonality property:
after integration over the unitary group, the integration measure $\int \d M$
reduces to an integration over the eigenvalues of $M$, and the Jacobian of
this transformation is a square Vandermonde determinant:
$$ P_n(\l) \propto\int \prod_{i=1,\ldots,n} \d \l_i\, \ee{-N
V(\l_i)}\left(\lambda-\lambda_i\right)  \,
\Delta^2(\l_1,\ldots,\l_n),$$
where
$$\Delta(\l_1,\dots \l_n)=\prod_{i<j} (\l_i-\l_j).$$
In this form we recognize a more classical expression \ref\rSze{G.
Szeg\"o, {\it Orthogonal Polynomials} ed. by the American Mathematical
Society, (1939).}. Then,
setting $\l=\l_0$:
$$ <P_n(\l_0) \cdot \l_0^m>\propto\int \prod_{i=0,\ldots, n} \d\l_i\, \ee{-N
V(\l_i)} \, \Delta(\l_0,\l_1,\ldots ,\l_n)\, \Delta(\l_1,\ldots ,\l_n) \l_0^m.
$$
The first Vandermonde is completely antisymmetric in the $n+1$ variables, we
can therefore antisymmetrize
the factor $\Delta(\l_1,\ldots ,\l_n) \l_0^m$, the result is zero if $m<n$
because the only polynomial completely antisymmetric and of degree less than
$n-1$ in $\l_0$ is zero. Thus
$$ <P_n \cdot \l^m> =0 \,.$$
it proves that $P_n$ is orthogonal to any $P_m$ with $m<n$, and that
$<P_n.P_m>=0$ as
soon as $m\neq n$.

Remark that a similar integral representation can also be found for
multi-matrix models. Consider two families of orthogonal polynomials $P_n$ and
$Q_n$, such that:
$$<P_n(\l) \cdot Q_m(\mu)>=\int\int \d\l\,\d\mu\, \ee{-N
(V(\l)+U(\mu)-c\l\mu)} P_n(\l) Q_m(\mu)\propto\delta_{nm}. $$
Then, we have:
$$P_n(\l)=\int\int \d M_1 \, \d M_2\, \ee{-N\tr[ V(M_1)+U(M_2)-cM_1 M_2]}
\det(\l-M_1)$$
where $M_1$ and $M_2$ are hermitian $n\times n$.
Similarly:
$$Q_m(\mu)=\int\int \d M_1 \, \d M_2\, \ee{-N\tr[V(M_1)+U(M_2)-cM_1 M_2]}
\det(\mu-M_2).$$

%----------------------------------------------------------------%
\appendix{2}{Connected correlation functions and the kernel
$\kappa(\l,\mu)$}

Some exact expressions exist for the correlation functions of matrix models
\ref\rDyson{F.J. Dyson, {\it Commun. Math. Phys.} 19(1970) 235; {\it
Correlations between the eigenvalues of a random matrix}.},
in terms of a kernel $\kappa(\l,\mu)$.
We again consider the matrix distribution \eprob:
$$  {\cal P}(M)={1\over Z} \ee{-N\tr V(M)}.$$
The corresponding measure can be rewritten in terms of the eigenvalues
$\lambda_i$ of $M$ and a unitary transformation $U$ which diagonalizes $M$:
$$  {\cal P}(M)\d M=Z^{-1}\d U \prod_{i=1\ldots N} \d \l_i\, \ee{-N V(\l_i)}
\, \Delta^2(\l_1, \ldots,\l_N)$$
($\Delta=\prod_{i<j} (\l_i-\l_j)$ being the Vandermonde determinant).
Therefore, the probability that the eigenvalues of $M$ are $\l_1,\dots,\l_N$
is:
$$\rho_N(\l_1,\ldots ,\l_N)\propto \prod_{i=1\ldots N} \ee{-N V(\l_i)} \,
\Delta^2(\l_1,\ldots ,\l_N),$$
and the correlation functions are obtained by partially integrating over some
eigenvalues:
$$\eqalign{\rho(\l_1)&=\int \prod_{i=2,\ldots, N} \d \l_i\, \rho_N(\l_1,\ldots
,\l_N) ,\cr\rho_2(\l_1,\l_2)&=\int \prod_{i=3,\ldots, N} \d \l_i\,
\rho_N(\l_1,\ldots ,\l_N)\cr}$$
... and so on.

The Vandermonde determinant $\Delta$ can be written as:
$$ \Delta(\l_i)=\det \l_i^{j-1},$$
and thus, after some linear combinations of columns of the matrix:
$$ \Delta=\det \Pi_{j-1}(\l_i),\qquad
\Pi_n(\lambda)=\lambda^n+O\left(\lambda^{n-1}\right),  $$
identity true for any set of polynomials $\{ \Pi_n \}$ normalized as above. In
order
to
perform the $\lambda$ integrations, we choose $\Pi_n\propto P_n$, $P_n$
being the orthogonal polynomials \ePns. $\Delta^2$ is the product of two such
determinants, therefore it is the determinant of a matrix product:
\eqna\eKDel
$$\eqalignno{\Delta^2 &\propto \det K(\l_i,\l_j) & \eKDel{a}\cr
K(\l_i,\l_j)&=\sum_{k=0}^{N-1} P_k(\l_i) P_k(\l_j).&\eKDel{b}\cr}$$
The proportionality constant in  \eKDel{a} is here irrelevant because the
eigenvalue distribution is normalized. The Darboux--Christoffel formula
(appendix 3)
tells
that:
$$K(\l,\mu)=\alpha{P_N(\l)P_{N-1}(\mu)-P_N(\mu)P_{N-1}(\l)\over \l-\mu}$$
where $\alpha$ is a normalization constant depending on $N$, and
$\alpha=(a-b)/4$ when $N\to\infty$.
The important properties of $K(\l,\mu)$ are:
$$\eqalign{\int\d\nu\, \ee{-NV(\nu)} K(\l,\nu)&=1\,,\quad
\int\d\nu\, \ee{-NV(\nu)} K(\nu,\nu)=N\,,\cr
\int \d\nu\, \ee{-NV(\nu)}K(\l,\nu)K(\nu,\mu)&=K(\l,\mu).\cr}$$
An explicit expression for $\rho_n$
$$\rho_n(\l_1,\dots,\l_n)={1\over N!}\int \prod_{i=n+1,\dots, N} \d\l_i\,
\prod_{i=1,\ldots, N}\ee{-N V(\l_i)} \,\det  K(\l_i,\l_j),$$
can be obtained by successively integrating over eigenvalues.
Using the rules of $K$ integration it is easy to prove by induction
$$\rho_n\left(\lambda_1,\lambda_2,\ldots,\lambda_n\right)=
{N^n(N-n)!\over N!}\det \kappa\left(\lambda_i,\lambda_j\right),$$
where we have introduced the reduced function
$$\kappa(\lambda,\mu)={1\over
N}\ee{-(N/2)[V(\lambda)+V(\mu)]}K(\lambda,\mu)\,.$$
Therefore, when the $\l_i$ are all distinct we have:
$$\left< \O(\l_1)\ldots \O(\l_n)\right>=\det
\kappa(\l_i,\l_j)=\sum_\sigma (-1)^\sigma \prod_i \kappa(\l_i,\l_{\sigma_i})$$
The connected function will involve only the sum over the permutations $\sigma$
such that $\prod_i \kappa(\l_i,\l_{\sigma_i})$ cannot be split in the product
of two cyclic products, i.e. only cyclic permutations will contribute to the
connected function. This intuitive result is a classical combinatorial
identity. %It
can be proven from%
%a simple identity (obtained for instance by fermion integration%
%methods),%
%$$\det \left[\delta_{ij}+s_i\kappa(\lambda_i,\lambda_j)\right]=%
%\sum_n{1\over n!}\sum_{i_1,i_2,\ldots,i_n}s_{i_1}s_{i_2}\ldots%
%s_{i_n}\det{}^{(n)}\kappa\left(\lambda_{i_l}\lambda_{i_k}\right),$$%
%Taking the logarithm of both sizes, and expanding in power series of the%
%%%sources%
%$s_i$, one obtains %
The connected function can thus be written:
$$\eqalign{\left< \O(\l_1)\ldots \O(\l_n)\right>_{\rm
conn}&=(-1)^{n+1}{1\over n} \sum_{{\rm permutations}\ \sigma} \prod_i
\kappa(\lambda_{\sigma_i},\l_{\sigma_{i+1}}),\cr
&=(-1)^{n+1}\left[\kappa\left(\lambda_1,\lambda_2\right)
\kappa\left(\lambda_2,\lambda_3\right)\ldots
\kappa\left(\lambda_{n},\lambda_1\right)+\cdots \right] ,\cr}$$
where the additional terms in the r.h.s.\ symmetrize the expression.
Or in a more compact way:
$$ \left< \O(\l_1)\ldots \O(\l_n)\right>_{\rm
conn}=(-1)^{n+1}
\sum_{{\rm cyclic\, permutations}\
\sigma\,}\,\prod_{i=1}^n \kappa\left(\l_i,\lambda_{\sigma_i}\right)$$
%--------------------------------------------------------%
\appendix{3}{Derivation of the Darboux--Christoffel formula}

The polynomial $\l P_n(\l)$ can be expanded on the basis of the $P_m$ with
$m\leq n+1$:
$$\l P_n(\l)= \sum_{m=n-1}^{n+1}Q_{nm} P_m(\l).$$
The orthogonality condition \ePns\ implies that the matrix $Q$ is symmetric:
$$Q_{nm}=<\l P_n \cdot P_m>=<P_n \cdot \l P_m>=Q_{mn}\,.$$
The polynomial $(\l-\mu)K(\l,\mu)$ can thus be written:
$$(\l-\mu)K(\l,\mu)=\sum_{k=0}^{N-1}\left(
\sum_{i=0}^N Q_{ki}P_i(\l)P_k(\mu)-\sum_{j=0}^N Q_{kj}P_k(\l)P_j(\mu)
\right).$$
All the terms cancel, except the upper-bounds:
$$(\l-\mu)K(\l,\mu)=Q_{N,N-1} \bigl( P_N(\l)P_{N-1}(\mu)-P_N(\mu)P_{N-1}(\l)
\bigr),$$
therefore $\alpha=Q_{N,N-1}$.
In the large $N$ limit, it is possible to calculate $\alpha$. The simplest way
of doing this is to calculate $\l P_{N-1}$ from expression \ePnasym. Since
$$\left( {a+b\over2}-{b-a\over2}\cos\varphi\right)\cos{(\psi-\varphi)}=
{a+b\over2}\cos{(\psi-\varphi)}-{b-a\over
4}\left(\cos\psi+\cos{(\psi-2\varphi)}\right)
,$$
we have:
$$\l P_{N-1}={a+b\over 2}P_{N-1}-{b-a\over4}\left( P_N+P_{N-2}\right)$$
and therefore $\alpha=(a-b)/4$.
\listrefs
\bye